\documentclass[reprint,amsmath,amssymb, aps,showpacs]{revtex4-1}

\usepackage{dcolumn}
\usepackage{bm}

\usepackage{graphicx}

\usepackage{here}
\usepackage{subfigure}

\begin{document}

\preprint{\bf PREPRINT}

\title{Characteristic power spectrum of the diffusive interface dynamics
in the two-dimensional Ising model }

\author{Yusuke Masumoto}
 \email{masumoto@scphys.kyoto-u.ac.jp}
\author{Shinji Takesue}
 \email{takesue@scphys.kyoto-u.ac.jp}
\affiliation{
Department of Physics, Kyoto University, Kyoto 6068502, Japan\\}
\date{\today}

\begin{abstract}
We investigate properties of the diffusive motion of an interface in the two-dimensional 
Ising model in equilibrium or nonequilibrium situations. We focused on the relation 
between the power spectrum of a time sequence of spins and diffusive motion of an 
interface which was already clarified in one-dimensional systems with a nonequilibrium 
phase transition like the asymmetric simple exclusion process. 
It is clarified that the interface motion is a diffusion process with a drift force toward
the higher-temperature side when the system is in contact with heat reservoirs
at different temperatures and heat transfers through the system.  Effects of the width of
the interface are also discussed.
\end{abstract}

\maketitle

\section{Introduction}
\begin{figure}
\centering 
\includegraphics[width=8.6cm]{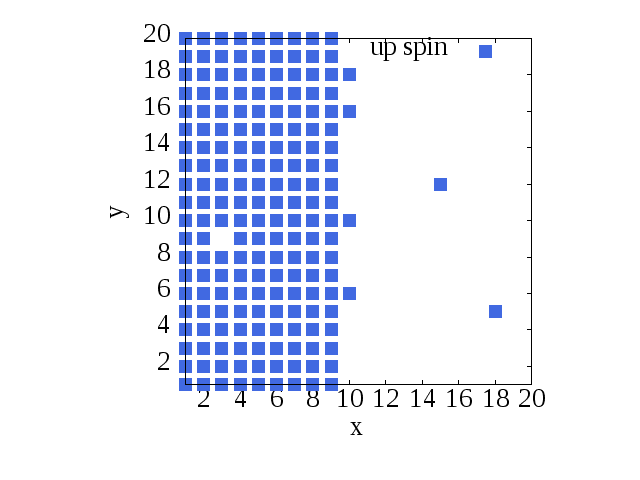} 
 \caption{Snapshot of an interface of the two-dimensional Ising model of
 size $20 \times 20$ at the temperature $T=1.5$. 
Sites with up spins are filled and those with down spins are blank.}
\label{f1}
\end{figure}

Dynamical and statistical properties of an interface between two phases
have been studied in relation to various physical phenomena 
such as phase ordering, crystal or dendrite growth, 
and self-propelled droplet motion driven by surface tension gradient\cite{11}. 
A simple example is found in the two-dimensional Ising model with 
ferromagnetic interaction. Below the critical temperature , if a magnetic field 
in opposite directions is imposed at the left and right boundaries, 
there appears an interface in the vertical direction as shown in Fig.~\ref{f1}.
If no magnetic field is applied in the bulk,
the interface carries out diffusive motion, while the bulk magnetic field drives 
the inter face motion in one direction, which is described by the KPZ equation
\cite{1}.

Actually diffusive motion of an interface is also seen in one-dimensional 
systems \cite{2}. Though no phase transitions occur in equilibrium 
one-dimensional systems, noneqilibrium models like the asymmetric 
simple exclusion process (ASEP) have the first-order phase transition 
where two phases coexist and an interface appears between them. 
One of the authors has clarified that characteristics of the diffusive 
motion of the interface are captured by the power spectrum that shows 
characteristic power law behavior with exponent $-3/2$ 
and the prefactor is determined by the diffusion constant, the difference 
of the density of two phases and the system size \cite{3}.

In this paper, we apply the method to the two-dimensional Ising model 
to investigate the interface motion. In particular, we discuss the relation 
between the interfacial width and the power spectrum. 
Moreover, when the boundaries of the systems are in contact with heat 
reservoirs at different temperatures, we find that a drift force towards the higher
 temperature side is generated accompanying heat conduction.  As the result,
 the probability of finding the interface is larger in the higher temperature 
region as reported by Yukawa {\it et al.} \cite{4}.

To investigate such noneqilibrium systems, use of Glauber dynamics is not 
appropriate because it needs a prescribed temperature value. 
In those cases deterministic energy-conserving dynamics such as Creutz 
and Q2R dynamics have been used \cite{4,5}.  At low temperatures,
however, those dynamics freezes and the interface 
hardly moves for a long time.  Thus, we need to devise a new dynamics 
to recover diffusivity in low temperature. Such dynamics is introduce in Sec.~I\hspace{-.1em}I . 
We apply the new dynamics to simulate heat conduction and analyze 
the power spectrum. We also compare it with other time evolution rules.

The remainder of the paper is organized as follows.  In Sec.~I\hspace{-.1em}I\hspace{-.1em}I, simulation 
results on heat conduction with and without an interface are exhibited. 
In either case, the thermal conductivity is estimated. Section I\hspace{-.1em}V is the main 
part of the paper, where we review the relation between the power law 
in the power spectrum and diffusive motion of the interface and 
apply it to the Ising model in equilibrium or noneqilibrium steady states. 
Section V is devoted to summary and conclusion.  

\section{Dynamics}
Because the Ising model does not have its own dynamics, we have to introduce 
some time evolution rule.
Glauber dynamics \cite{6} is most commonly used for simulation of 
equilibrium states. 
In Glauber dynamics, temperature appears as a parameter and the realization 
of the equilibrium state at the temperature is guaranteed.  In heat conduction, 
however, temperature values are given only at the boundaries and the bulk 
temperature is determined as the result of dynamics. Thus, the Glauber 
dynamics is not suitable for the simulation of heat conduction.

\begin{figure}
\centering 
\includegraphics[width=8.6cm]{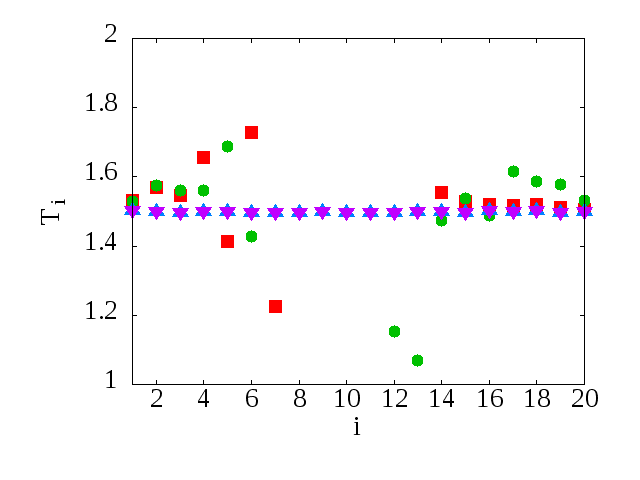} 
\caption{Temperature profiles generated by Creutz dynamics and KSC dynamics. 
The system size is $20\times20$ and the temperature of heat reservoirs
 is $T_R=T_L=1.5$. Red squares (green circles) indicate 
the result for the system with (without) 
an interface 
obtained by average over $9\times10^7$ steps after $10^7$ steps of transients. 
Blue upward triangles (purple downward triangles) 
indicate the result for the system with (without) an interface obtained by 
KSC dynamics calculated in the similar way as Creutz dynamics. }
\label{f2}
\end{figure}
\begin{figure*}[thbp]
\begin{center}
\subfigure[]{
\includegraphics[width=86mm]
{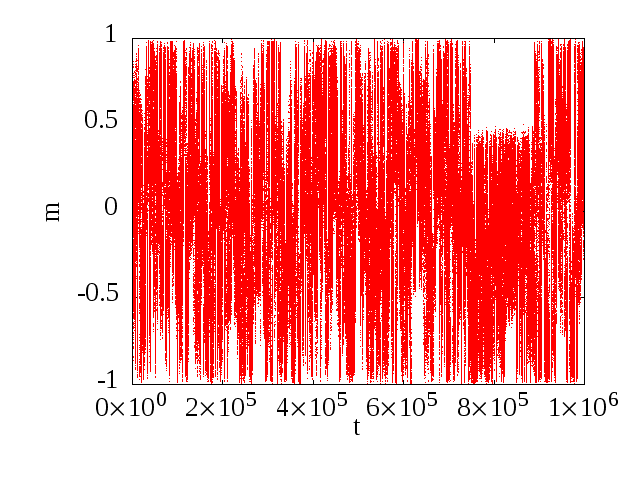} 
\label{f3}}
\subfigure[]{
\includegraphics[width=86mm]
{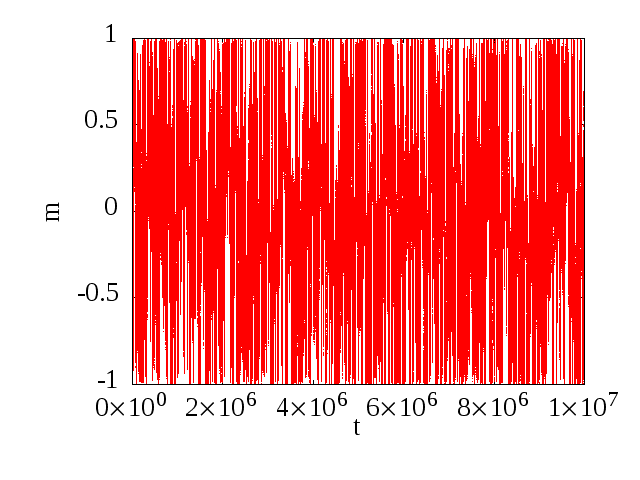}
\label{f3b}}
\end{center}
\caption{(a) Evolution of magnetization for $10^6$ time steps using Creutz dynamics 
in the system with an interface. System size is $20\times20$ and the temperature of heat reservoirs
is $T_R=T_L=1.5$. (b) Evolution of magnetization by KSC dynamics in the system with interface
obtained in the similar system size and at the similar temperature as by Creutz dynamics.}
\label{f3and4}
\end{figure*}
Creutz devised an alternative dynamics for the Ising model. 
In the case of 
the  square lattice, it is a dynamics that conserves the following 
Hamiltonian \cite{5,7}
\begin{equation}
H=-\sum_{i,j}(\sigma_{i,j}\sigma_{i+1,j}+\sigma_{i,j}\sigma_{i,j+1})+\sum_{i,j}4\tilde{\sigma}_{i,j}, 
\label{eq2}
\end{equation}
where $\sigma_{i,j}\in\{-1,+1\}$ denotes the Ising spin on site $(i,j)$ 
and $\tilde{\sigma}_{i,j}\in\{0,1,2,3\}$ is an auxiliary variable 
called ``momentum.''  The first term means the usual ferromagnetic 
interaction and the second term is a kind of ``kinetic energy.'' 
In each step, spin $\sigma_{i,j}$ flips if and only if the change of the 
interaction energy can be compensated by some change of momentum 
variable $\tilde{\sigma}_{i,j}$. The condition is given by
\begin{equation}
0\leq\tilde{\sigma}_{i,j}-\frac{1}{2}\sigma_{i,j}(\sigma_{i-1,j}+\sigma_{i+1,j}+\sigma_{i,j-1}+\sigma_{i,j+1})\leq3  \label{eq3}
\end{equation}
and when it is satisfied,  $\sigma_{i,j}$ changes its sign and 
$\tilde{\sigma}_{i,j}$ becomes 
${\tilde{\sigma }_{i,j}}'=\tilde{\sigma}_{i,j}
-\frac{1}{2}\sigma_{i,j}(\sigma_{i-1,j}+\sigma_{i+1,j}
+\sigma_{i,j-1}+\sigma_{i,j+1})$. 
If we divide the lattice into two sublattices just as white and black fields of 
the chessboard, the Creutz dynamics can be carried out simultaneously 
for spins on a sublattice chosen alternatingly. Then, the dynamics is fully 
deterministic. An alternative way is to choose a site randomly and examine 
if its spin can be flipped. In the following the latter is employed.
A simplified variant of Creutz dynamics is Q2R, 
where the ``kinetic energy''term is absent and spins can flip only if the sum 
of the four nearest-neighbor spins is zero.

In an isolated system, the total energy is conserved under Creutz or Q2R 
dynamics and the system reaches equilibrium in a long run, 
provided that the dynamics is sufficiently ergodic. 
Attachment of heat reservoirs to parts of the system is straightforward. 
We only have to make spins in contact with heat reservoirs 
evolve according to Glauber dynamics. 
The temperature of the left heat reservoir is $T_L$ and that of the right
 heat reservoir is $T_R(\leq T_L)$. 
Note that we employ an energy unit where Boltzmann constant
is unity. Thus, the critical temperature of the two-dimensional Ising model is
$T_c=2/\log(1+\sqrt{2})$.
Thus, we can simulate heat conduction using such 
dynamics\cite{5}. Moreover, if both $T_L$ and $T_R$ are below
$T_c$ and  $+$ spin is fixed at the boundary of the left reservoir
 and $-$ spin is fixed at the boundary of the right reservoir, 
an interface is formed.
Because momentum variables are expected to obey the canonical distribution in local 
equilibrium states, local temperature can be estimated 
by measuring the momentum distribution.

At low temperature, however, Creutz and Q2R dynamics share a common problem. 
Below the critical temperature, most spins orient in the same direction. 
Thus, large momentum is necessary to compensate energy increase brought 
by a spin flip. However, there are few large momentum in low temperature 
and accordingly the dynamics freezes.  Figure \ref{f2} shows the temperature profile 
after large time steps by Creutz dynamics with two heat reservoirs, 
both at temperature $1.5$, 
where the system does not  reach equilibrium at uniform temperature 
in simulation time. 
In \cite{5}, it is reported that the thermal conductivity drops abruptly below 
the critical temperature in the case of no fixed spins at the boundaries. 
When an interface is  present, the magnetization is stuck and 
the diffusive motion of an interface is subdued as seen in Fig.~\ref{f3}.

A solution to this problem was brought by Casartelli {\it et al}. \cite{8}. 
They studied Q2R and noticed the same problem. Then they modified Q2R 
by adding a new rule of spin flip to it. It is called Kadanoff-Swift (KS) dynamics, 
where a pair of next nearest-neighbor spins exchange their values 
if energy is unchanged by the exchange. It should be noted that the Hamiltonian 
does not include next-nearest-neighbor coupling. Such spins are only 
dynamically coupled. To measure local temperature in nonequilibrium systems, 
we apply the same modification to Creutz dynamics and we call the resultant 
dynamics Kadanoff-Swift-Creutz (KSC) dynamics.

Precisely, the algorithm that we employ for our KSC dynamics is the following.\\

\begin{enumerate}
 \item Choose random Creutz dynamics or KS dynamics.
 \item In the former case, a site is randomly chosen and its spin is updated 
according to Creutz dynamics. In the latter case, a couple of nearest-neighbor sites
are randomly chosen and the spins are updated according to KS dynamics.
 \item Repeat 1 and 2 $L_xL_y$ times, where $L_x$ and $L_y$ are 
the horizontal and vertical size of the system.
 \item Update the spins in contact with reservoirs according to Glauber dynamics.
 \item Repeat the procedure from 1 to 4, which is counted as a unit of time.
\end{enumerate}

In this dynamics,  
relaxation to an equilibrium state at uniform temperature is realized as seen in 
Fig.~\ref{f2} even at a 
temperature considerably below the critical temperature.
Moreover, Fig.~\ref{f3b} shows that 
the magnetization changes smoothly from $-1$ to $1$.

\section{Thermal conductivity}

\begin{figure}[htbp]
\centering 
\setlength\textfloatsep{0pt} 
\includegraphics[width=8.6cm]{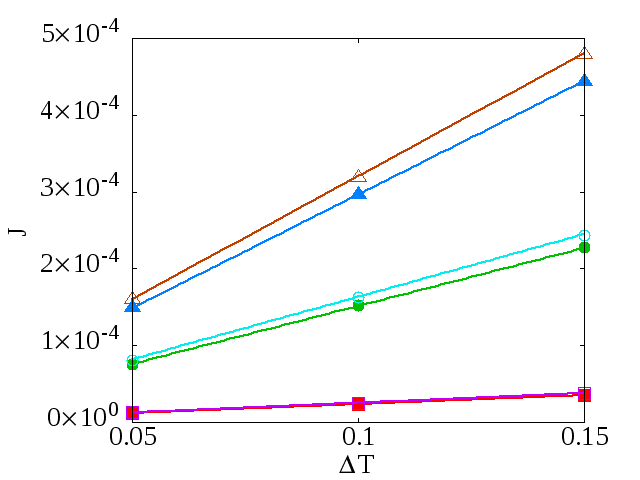}
 \caption{Heat flux in the system with and without an interface. 
The system size is $40\times40$ and the average temperature of heat reservoirs 
is $T_{\mathrm{ave}}=1.0,1.5,2.0$. 
Data is obtained from averaging over 50 samples for $4.5\times10^7$ steps each. 
Red (filled squares), green (filled circles) and blue dots (filled triangles) indicate the results of $T_{\mathrm{ave}}=1.0$, $1.5$, 
and $2.0$, respectively, in the system with an interface and light red (open squares), light blue (open circles), and 
yellow dots (open triangles) indicate those of $T_{\mathrm{ave}}=1.0,1.5$, and $2.0$, respectively,
in the system without an interface.}\label{f4}
\end{figure}

First we examine how the new dynamics and the existence of 
an interface affect Fourier's law and the thermal conductivity. 
We have carried out numerical simulation for 
the system of size $40\times40$, both horizontal ends of which are in contact 
with heat reservoirs at different temperatures with difference $\Delta T=T_L-T_R$. 
The time evolution of the system is governed by the KSC dynamics. 
Then the temperature gradient is formed in the system as in the Creutz 
dynamics \cite{5,7}. As we see in Fig.~\ref{f4}, heat flux $J$ 
is proportional to the temperature difference $\Delta T$ irrespective of the 
existence or nonexistence of an interface.

\begin{figure}[htbp]
\centering 
\includegraphics[width=8.6cm]{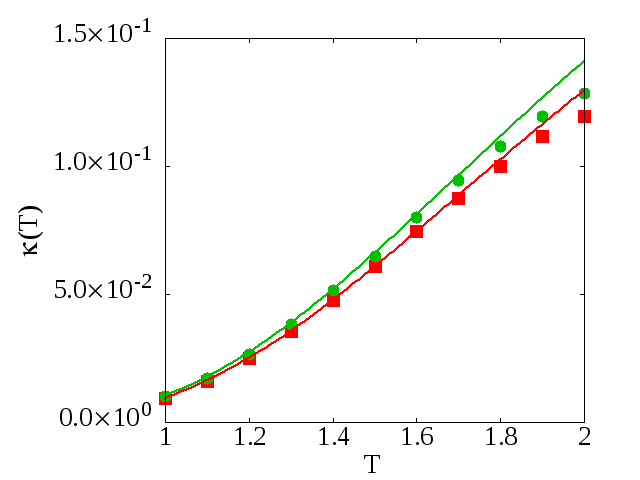}
 \caption{Thermal conductivity $\kappa(T)$ by KSC dynamics. 
Red squares and green circles indicate the results of the system with  and without 
an interface respectively. 
Red (lower) and green (upper) lines indicate $28.3/T^2\exp(-8/T)$ and $30.8/T^2\exp(-8/T)$ 
respectively.}\label{f5}
\end{figure}

However, values of the thermal conductivity show difference between the two cases. 
The thermal conductivity is defined as 
\begin{equation}
\kappa(T)=J\frac{L_x}{\Delta T}.  \label{eq5}
\end{equation}
Figure \ref{f5} shows the thermal conductivity in the system 
with KSC dynamics. 
In low temperature, the thermal conductivity varies 
like  $\kappa\sim \frac{1}{T^2}\exp(-\frac{8}{T})$  irrespective of the presence or
 absence of an interface, as was suggested in \cite{5}. 
In the case of the KSC dynamics,
the thermal conductivity is larger in the system without an interface 
than in that with an interface. 
Interestingly, the converse is the case when the Creutz dynamics is employed.

\begin{figure*}[thbp]
\begin{center}
\subfigure[]{
\includegraphics[width=86mm]
{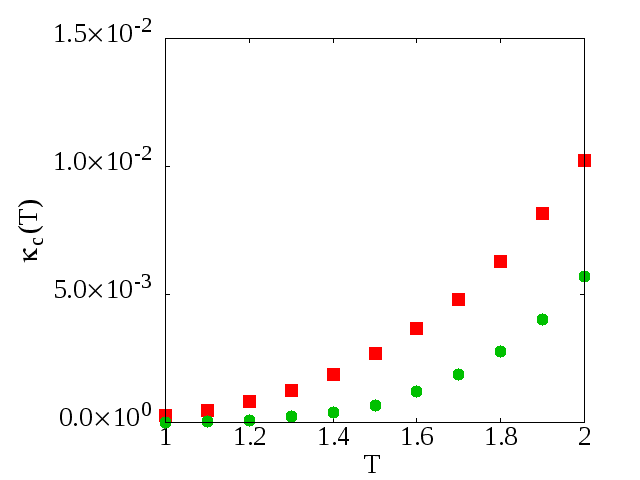} 
\label{f6}}
\subfigure[]{
\includegraphics[width=86mm]
{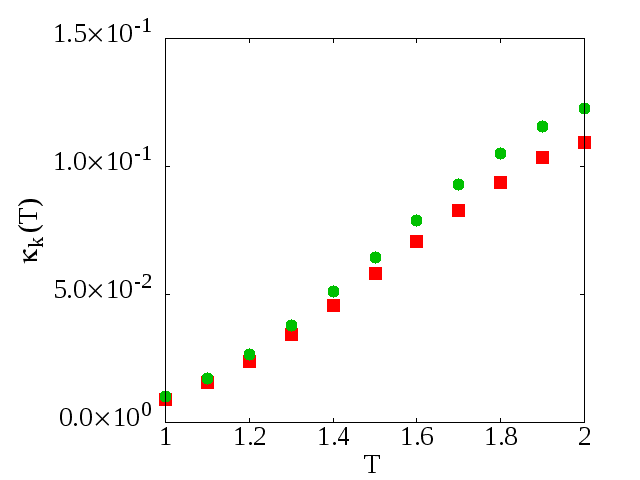}
\label{f7}}
\end{center}
\caption{
(a) Contribution from Creutz dynamics to the thermal conductivity, $\kappa_c(T)$. 
Red squares and green circles indicate the results of the system with and without an interface 
respectively. (b) Contribution from KS dynamics to the thermal conductivity, 
$\kappa_k(T)$. 
Red squares and green circles indicate the results of the system with and without an interface 
respectively.}
\label{f7and8}
\end{figure*}

\begin{figure}[thbp]
\begin{center}
\subfigure[]{
\includegraphics[width=40mm,height=50mm]
{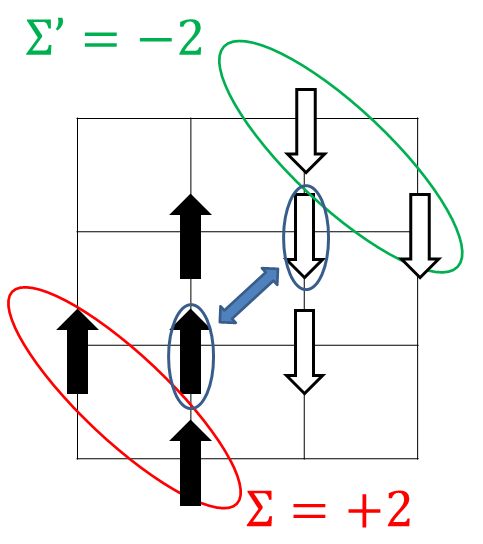} 
\label{f8}}
\subfigure[]{
\includegraphics[width=40mm,height=50mm]
{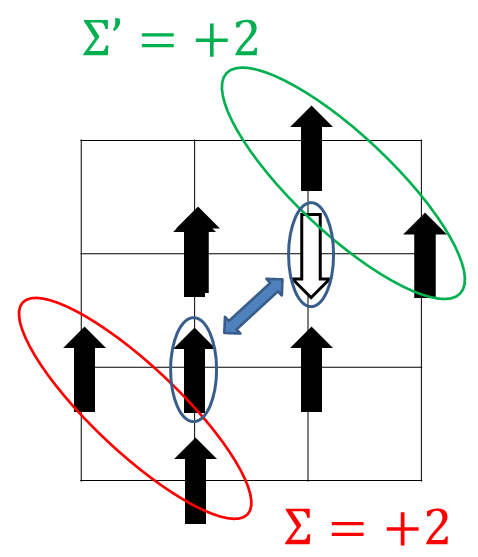}
\label{f9}}
\end{center}
\caption{Example of spin configuration. Panel (a) is the case where the two opposite spins are adjacent at an interface. Panel (b) is the case where the two opposite spins are not adjacent at an interface. }
\label{f8and9}
\end{figure}

Let us divide heat flux $J$ into contribution from Creutz dynamics $J_c$ and that
from KS dynamics $J_k$. We defined $\kappa_c$ and $\kappa_k$ as 
thermal conductivity computed using $J_c$ and $J_k$, respectively. 
As Figs.~\ref{f6} and \ref{f7} show,
$\kappa_c$ becomes larger while $\kappa_k$ becomes smaller when an interface 
is present.  The reason for this is explained as follows. 
The condition of flip in Creutz dynamics depends on change of spin-interaction
energy by the flip.  Because spins of different signs are adjacent at an interface,
a spin flip can occur with a small change of energy there.  If there is no interface,
a spin is likely to be surrounded by spins of the same sign and large energy is necessary
to flip.  Thus Creutz dynamics contributes more in conducting energy 
when an interface exists. 
On the contrary, KS dynamics is suppressed near an interface.
Consider a pair of next-nearest-neighbor spins at a flat interface as shown in Fig.~\ref{f8and9}.
If they exchange their signs, interaction energy with the spins marked $\Sigma$ and 
$\Sigma'$ changes and the total energy increases by 8.  Thus it is inhibited.
Apart from an interface, if only one spin is surrounded by opposite spins, it can exchange
its sign with a next-nearest-neighbor spin without change of energy.
Thus the presence of an interface controls the KS dynamics and makes $\kappa_k$
smaller.


\section{probability distribution of the interface position}
\begin{figure*}[thbp]
\begin{center}
\subfigure[]{
\includegraphics[width=86mm]
{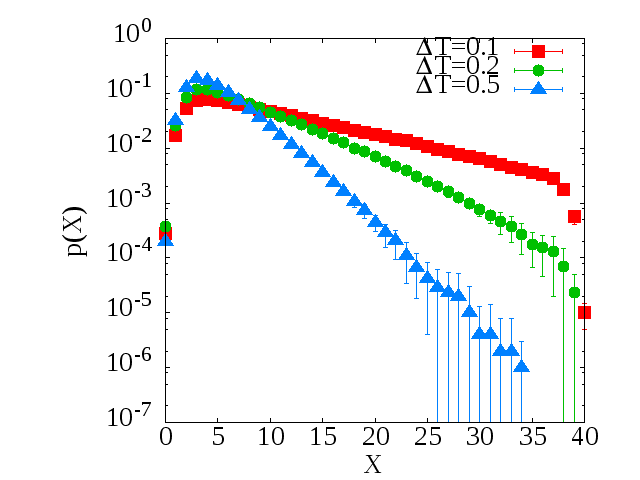} 
\label{f10}}
\subfigure[]{
\includegraphics[width=86mm]
{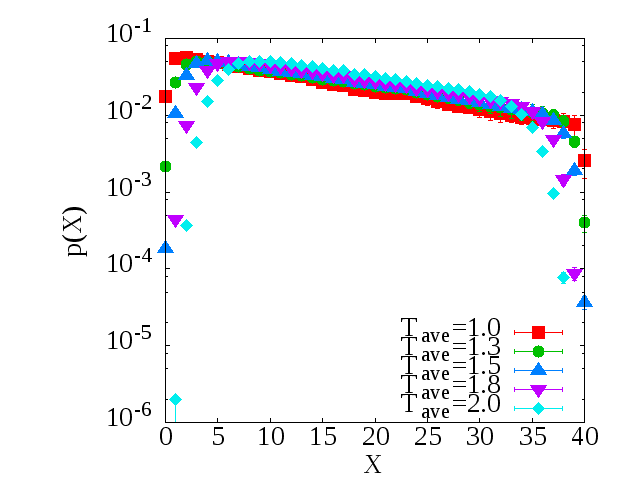}
\label{f11}}
\end{center}
\caption{(a) Probability distribution of the interface position in 
KSC dynamics with heat reservoirs whose average temperature is 
$T_{\mathrm{ave}}=1.5$ and $\Delta T=0.1, 0.2$, and $0.5$. 
(b) Probability distribution of the interface position when $\Delta T=0.05$ and
$T_{\mathrm{ave}}=1.0$, 1.3, 1.5, 1.8, and 2.0. }
\label{f10and11}
\end{figure*}

We define the interface position by using magnetization 
$m=(L_xL_y)^{-1}\sum_{i,j}\sigma_{i,j}$ as 
\begin{equation}
X=\frac{L_x}{2}(m+1). \label{eq6}
\end{equation}
Thus, the interface position is at the left edge $X=0$ if 
$m=-1$, at the right edge $X=L_x$ if $m=+1$, and in the center if $m=0$. 
We numerically estimate the probability distribution of the interface position in the 
system of size 40$\times$40 with KSC dynamics and heat reservoirs. 
Figure \ref{f10} is the result for $T_{\mathrm{ave}}=1.5$ and various $\Delta T$, 
which shows that the interface prefers to exist in the high temperature region.
The distribution is roughly exponential and the tendency is stronger with larger $\Delta T$. 
The result agrees with \cite{4}, where Creutz dynamics is employed, though. The exponential distribution is more evident when $\Delta T$ is small 
as seen in Fig.~\ref{f11}. 

We notice that the decay of $p(X)$ close to the boundary is faster for 
higher temperatures.  It is explained as follows.
The magnetization on the right (left) end is fixed to $+1$ ($-1$) by the 
boundary condition. Because its magnitude is larger than the spontaneous 
magnetization at temperature $T_L$ or $T_R$, 
the interface cannot approach the very ends.
The movable range of the interface is smaller than $[L'/2,L_x-L'/2]$,
where $L'$ denotes the width of the interface. 
Because the width of the interface is greater for higher temperatures,
the movable range is smaller. As is discussed later,
the change of the effective system size plays an role
in the temperature dependence of the power spectrum.

\begin{figure}[htbp]
\centering 
\includegraphics[width=8.6cm]{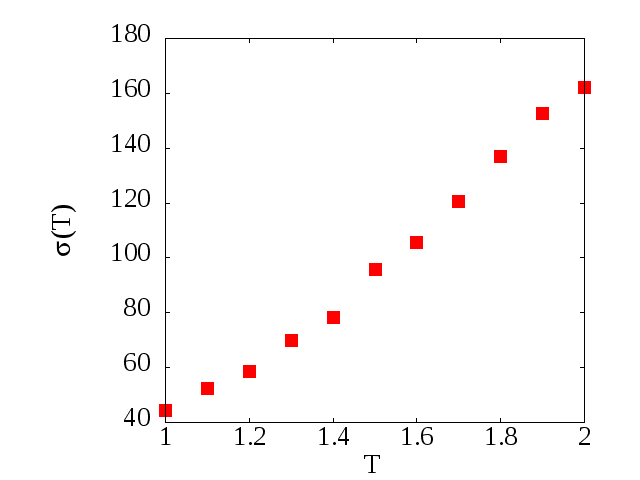}
 \caption{Interface energy estimated by using Eq.\/(\ref{eq10}) from the probability 
distribution.}\label{f12}
\end{figure}

Actually the exponential distribution is verified on 
the assumption of local equilibrium as follows.
The local equilibrium means that we can represent the probability distribution using 
temperature $T(x)$ and as  

\begin{equation}
p(X)\sim \exp\left(-\frac{\sigma(T,L_y)}{T(X)}\right), \label{eq7}
\end{equation}

where $\sigma(T,L_y)$ is the interface energy at $T$. 
If $\Delta T$ is very small, $1/T(x)$ can be approximated as
\begin{equation}
\frac{1}{T(x)}\sim \frac{1}{T_{\mathrm{ave}}}
\left(1+\frac{\Delta T}{T_{\mathrm{ave}}} \frac{x}{L} \right)\label{eq9} 
\end{equation}
and $\sigma(T,L_y)$ is regarded as a constant.  Then,
the probability distribution becomes

\begin{equation}
p(X)\sim \exp\left(-\sigma(T_{\mathrm{ave}},L_y)\frac{\Delta T}{{T_{\mathrm{ave}}}^2} 
\frac{X}{L}\right). \label{eq10}
\end{equation}

The interface energy $\sigma(T,L_y)$ thus estimated is shown  in Fig.~\ref{f12}.

\section{property of the power spectrum induced by the diffusive motion of interface in two dimensional Ising model.}
\subsection{Relationship between the power spectrum and the diffusive motion of the interface}

Let us briefly review the relation between the power spectrum and the diffusive 
motion of an interface according to \cite{3}.
Consider a one-dimensional system of size $L$, 
where the left part is in one phase of density $\rho_-$ and the right part is 
in the other phase of density $\rho_+(>\rho_-)$ and an interface 
at position $x(t)$ divides the two phases.
In this case, the density at site $y$ ($\rho_y(t)$) is described as
\begin{equation}
\rho_y(t)=\rho_{-}+(\rho_{+}-\rho_{-})\theta(y-x(t)), \label{eq11}
\end{equation}
where $\theta(x)$ is the step function.  
Now let us assume that $x(t)$ executes Brownian motion with diffusion constant $D$. 
Then, the probability density of finding the interface at $x$ obeys 
the diffusion equation.
\begin{equation}
\frac{\partial P}{\partial t}=D\frac{\partial^2 P}{\partial x^2}. \label{eq12}
\end{equation}
Under the reflecting boundary condition
\begin{equation}
\frac{\partial P(x,t)}{\partial x}=0 \label{eq13}
\end{equation}
at $x=0$ and $L$, 
the stationary state is $P_{st}(x)=L^{-1}$ and the transition probability is 
\begin{equation}
 P(x,t|x_0,t_0)
=\frac{1}{L}+\frac{2}{L}\sum_{n=1}^{\infty}e^{-D\lambda_n^2t}
\cos\lambda_nx\cos\lambda_n x_0,
\end{equation}
where $\lambda_n=\frac{n\pi}{L}$.
Then the autocorrelation function of 
$\delta\rho_y(t)=\rho_y(t)-\langle\rho_y\rangle_{st}$ is obtained as
\begin{equation}
 \langle\delta\rho_y(t)\delta\rho_y(0)\rangle=
\frac{2(\rho_+-\rho_-)^2}{L^2}\sum_{n=1}^{\infty}
\frac{e^{-D\lambda_n^2|t|}}{\lambda_n^2}\sin^2\lambda_ny.
\end{equation}
Thanks to the Wiener-Khinchin theorem, the power spectral density is given by the
Fourier transform of the correlation function, which is calculated as
\begin{eqnarray}
 I(\omega)&=&\int_{-\infty}^{\infty}e^{-i\omega t}\langle\delta\rho_y(t)
\delta\rho_y(0)\rangle dt
\nonumber \\
&=&\frac{2(\rho_+-\rho_-)^2}{L^2}
\sum_{n=1}^{\infty}\frac{2D}{D^2\lambda_n^4+\omega^2}
\sin^2\lambda_ny
\end{eqnarray}
If we assume $y=L/2$ and large $L$, the sum can be replaced by an integral which is
evaluated using the residue calculus.  Thus we arrive at
\begin{equation}
I(\omega)=(\rho_{+}-\rho_{-})^2\frac{\sqrt{2D}}{2L}\omega^{-1.5}. 
\label{eq14}
\end{equation}

To extend the above relation to the two-dimensional Ising model, 
we consider the following two types of sequences and their power spectrum. 
One is the power spectrum of a temporal sequence of column-averaged magnetization
\begin{equation}
\{s_{i}(t)=\frac{1}{L_y}\sum_j\sigma_{i,j}(t)|t=0,1,2,...,T-1\}. \label{eq15}
\end{equation} 
In our simulations, the length of the sequence is $T=2^{20}$. 
Fourier components of the sequence (\ref{eq15}) are computed as
\begin{equation}
S_n=\sum_{t=0}^{T-1}s_i(t)\exp(-\omega_n t)\label{eq16}
\end{equation}
where $\omega_n=\frac{2\pi n}{T}$.  
Then, the power spectrum $I(\omega)$ is derived as
\begin{equation}
I(\omega_n)=T\langle{|S_n|^2}\rangle, \label{eq17}
\end{equation}
where the brackets mean averaging over 200 samples. 
If the interface behaves as normal diffusion with diffusion constant $D$, 
we expect to obtain the power spectrum of the form (\ref{eq14})
with replacement of $\rho_+-\rho_-$ with $2m_0$,
where $m_0$ is the spontaneous magnetization\cite{9}
\begin{equation}
m_0(T)=[1-\sinh(2/T)^{-4}]^{1/8}. \label{eq19}
\end{equation}

The other is for a temporal sequence of spin values of a site
$\{\sigma_{i,j}(t)|t=0,1,2,\dots,T-1\}$.  
Its power spectrum is defined in the same manner as above. 

\subsection{Equilibrium System with Glauber dynamics}




\begin{figure*}[thbp]
\begin{center}
\subfigure[]{
\includegraphics[width=86mm]
{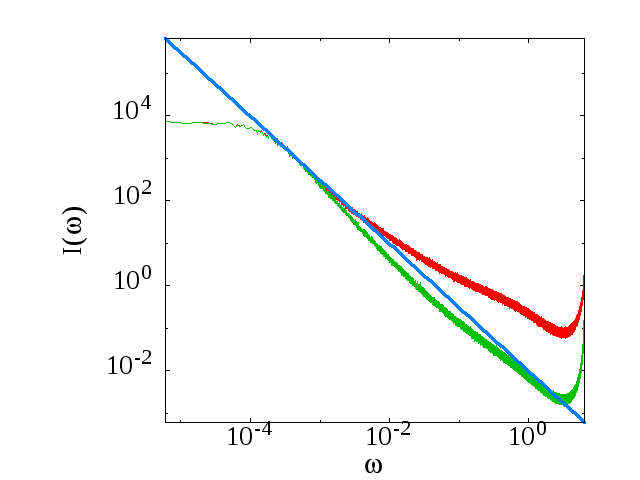} 
\label{f13}}
\subfigure[]{
\includegraphics[width=86mm]
{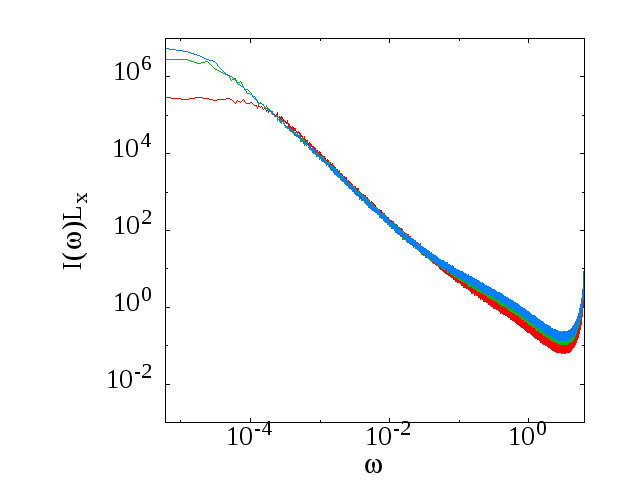}
\label{f14}}
\end{center}
\caption{(a) Power spectra for $\sigma_{L_x/2,L_y/2}(t)$ 
and $s_{L_x/2}(t)$ in the system of $L_x=L_y=40$ at $T=2.0$. 
Red (upper) dots indicates the former and green (lower) dots the latter. 
Blue line indicates $-1.5$ power law. (b) Plot of $I(\omega)L_x$ vs $\omega$ for $L_x=40$ [red (lower)],  80 [green (middle)], and 100 [blue (upper)] for the 
power spectrum of $s_{L_x/2}(t)$. 
The data collapse shows the $L_x$ dependence of Eq.~(\ref{eq17})}
\end{figure*}

\begin{figure}[htbp]
 \centering
   \includegraphics[width=86mm]{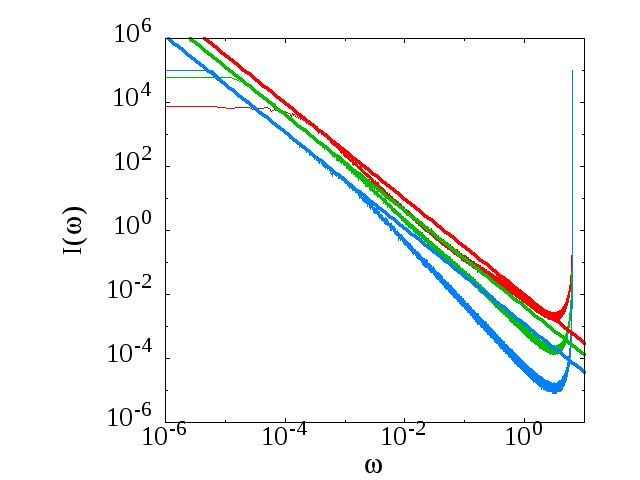}
 \caption{Temperature dependence of the power spectra of $s_{L_x/2}(t)$ ($L_x=L_y=40$).
Red (upper), green (middle), and blue (lower) dots indicate the results for $T=2.0$, $1.0$, and $0.5$, respectively.
Red (upper), green (middle), and blue (lower) lines indicate $-1.5$ power law at  $T=2.0$, $1.0$, and $0.5$, respectively.}
  \label{f15}
\end{figure}

First we examine the power spectra in the two-dimensional Ising model with 
Glauber dynamics. 
Figure \ref{f13} shows the power spectrum at $T=2.0$ slightly lower than 
the critical temperature $T_c$. 
As the figure shows, both the power spectra of the sequence of a spin 
on site ($\sigma_{L_x/2,L_y/2}(t)$) and the column-averaged magnetization 
at $L_x/2$ ($s_{L_x/2}(t)$) show power-law behavior
with exponent $-1.5$ in a range of low frequencies.   
Moreover, Fig.~\ref{f14} shows data collapse for
horizontal-size variation according to Eq.~(\ref{eq14}). These results indicate that
the interface undergoes normal diffusion in a certain time scale.  
Furthermore, temperature dependence of the power spectra shown in
Fig.~\ref{f15} indicates that the diffusion constant becomes smaller as temperature decreases.
\begin{figure}[htbp]
\centering
 \includegraphics[width=86mm]{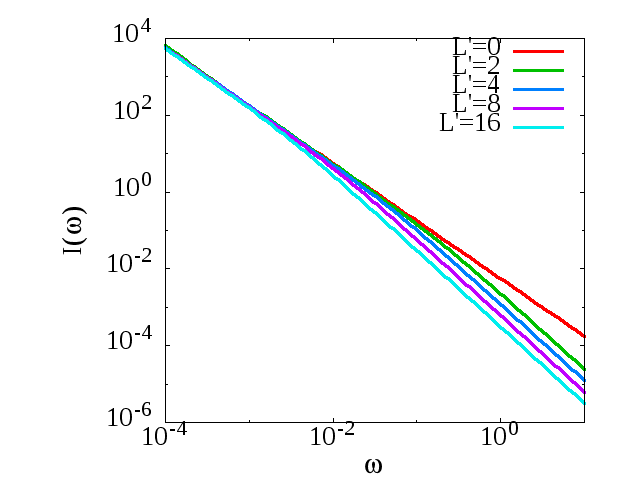}
\caption{Power spectra described by Eq.~(\ref{eq29}) for a variety of width $L'$.}
  \label{f16}
\end{figure}
\begin{figure}[htbp]
\centering
 \includegraphics[width=86mm]{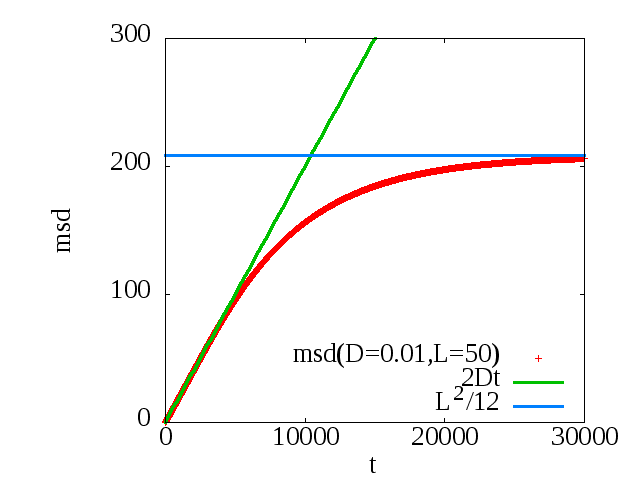}
\caption{Mean square displacement of Eq.(\ref{eq31}) ($L=0.01,L=50$).}
  \label{f17}
\end{figure}
\begin{figure}[htbp]
\centering
 \includegraphics[width=86mm]{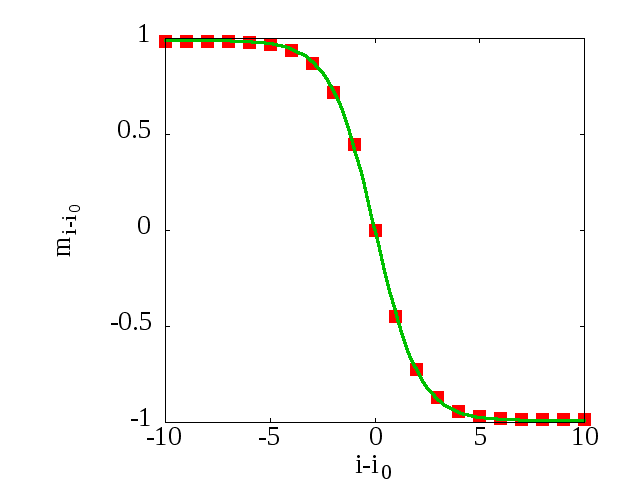}
\caption{Profile of column-averaged magnetization fitted by tanh profile Eq.~(\ref{eq33}). 
Red squares indicate the numerical result at $T=1.5$ and 
green line indicates $-0.99\tanh(2/4.22(i-i_0))$.}
  \label{f18}
\end{figure}
However, the power-law behavior is limited to a short range of low frequencies and the graph
shows a characteristic curve in a higher frequency region. 
In the case of the power spectrum of 
column-averaged magnetization $s_{L_x/2}(t)$, the deviation is explained by taking into account
 width of the interface.  In the one-dimensional case, an instantaneous density profile is a 
step function without a width.  Contrastingly, the column-averaged magnetization has a width.  
Thus we have to consider its influence on the spectrum.
In place of the step-function profile, 
let us assume that the $s_x(t)$ is approximately described as

\begin{equation}
s_x(t)=
  \begin{cases}
    m_0 &(0 \leq x <{X(t)-\frac{L'}{2}})  \\
    \frac{2m_0}{L'}(x-X(t)) & ({X(t)-\frac{L'}{2}} \leq x < {X(t)+\frac{L'}{2}})\\
    -m_0 &({X(t)+\frac{L'}{2}} \leq x \leq L) 
 \label{eq20}   
  \end{cases}
\end{equation}
where $L'$ denotes the width and $X(t)$ is the position of the interface at time $t$, which 
undergoes Brownian motion with diffusion constant $D$.
In this case the autocorrelation function of $\delta s_y(t)=s_y(t)-\langle s_y\rangle_{st}$ is 
calculated as 
\begin{equation}
\langle\delta s_y(t)\delta s_y(0)\rangle 
= \frac{8m_0^2}{L^2}\sum_{n}e^{-D\lambda_n^2|t|}\frac{\sin^2(\frac{L'}{2}\lambda_n)}{(\frac{L'}{2}\lambda_n)^2}\frac{\sin^2{\lambda_ny}}{\lambda_n^2},
\label{eq26}
\end{equation} 
where $\lambda_n=n\pi/L$, and the power spectral density is its Fourier transform
\begin{eqnarray}
I(\omega)&=&\int_{-\infty}^{\infty} \exp(-i\omega t)\langle{\delta s_y(t)\delta s_y(0)}
\rangle dt \nonumber\\
         &=&
\frac{16Dm_0^2}{L^2}\sum_{n=1}^{\infty}\frac{\sin^2(\frac{L'}{2}\lambda_n)}{(\frac{L'}{2}\lambda_n)^2}\frac{\sin^2{\lambda_ny}}{D^2\lambda_n^4+\omega^2}.
\label{eq27}
\end{eqnarray}
Putting $y=L/2$ and  assuming that $L$ is large, 
the sum over $n$ can be replaced by the following integral                             
\begin{equation}
I_E(\omega)\sim\frac{32m_0^2}{\pi DL}\int_{0}^{\infty}\left(
\frac{\sin\beta x}{\beta x}\right)^2\frac{dx}{x^4+\alpha^2},
 \label{eq28}
\end{equation}
where $\alpha=\frac{\omega}{D}$ and $\beta=\frac{L'}{2}$.

This integral is evaluated using residue calculus and we arrive at
\begin{widetext}
\begin{equation}
I_E(\omega)=(2m_0)^2\left[\frac{2D^{\frac{3}{2}}}{LL'^{2}}\omega^{-\frac{5}{2}}
\left\{\exp(-L'(\frac{\omega}{2D})^\frac{1}{2})
\cos(L'(\frac{\omega}{2D})^\frac{1}{2}+\frac{\pi}{4})-\frac{1}{\sqrt{2}}\right\}
+\frac{2D}{LL'}\omega^{-2}\right].
\label{eq29}
\end{equation}
\end{widetext}

Figure \ref{f16} shows $I_E(\omega)$ for various values of width $L'$.
They obey the power law with exponent $-1.5$ in 
a low-frequency region but the slope becomes
steeper beyond a certain value that is a decreasing function of the width.

To compare Eq.~(\ref{eq29}) with the numerically obtained power spectra, 
we need the diffusion constant $D$ and the width of the interface $L'$. 
Moreover, because the boundary condition is different,  we treat the system size $L$ 
as a fitting parameter.

The diffusion constant and the effective system size are numerically estimated as follows.
We prepare a system with a straight interface at $x=L/2$ and measure the evolution of 
the mean square displacement (msd), which is analytically derived as 
\begin{equation}
\langle(X-L/2)^2\rangle
=\sum_{k=1}\frac{L^2}{\pi k^2}\{\exp(-\frac{4\pi^2 Dk^2}{L^2}t)-1\}(-1)^k \label{eq31}
\end{equation}
As Fig.~\ref{f17} shows, the msd behaves like $2Dt$ for small $t$ and 
converges to $L^2/12$ for large enough $t$.   
The numerical results thus obtained indicate that $L \simeq L_x$ at small temperature, 
while $L$ becomes smaller than $L_x$ at higher temperature.

The interfacial width $L'$ is estimated by fitting the profile of column-averaged 
magnetization $s_i(t)$.  Given a profile $s_i(t)$ at time $t$, we determine the interface position
$i_0(t)$ as the position that gives minimum $|s_i(t)|$.  Then the time average of $s_{i-i_0(t)}(t)$
is well fitted by the function
\begin{equation}
-m_0\tanh(2i/L'), \label{eq33}
\end{equation}
as shown in Fig.~\ref{f18}.
For example, at $T=1.5$, $L'$ is estimated to be 4.22 as shown in Fig.~\ref{f18}.

\begin{figure*}[htbp]
\begin{center}
\subfigure[]{
\includegraphics[width=86mm]
{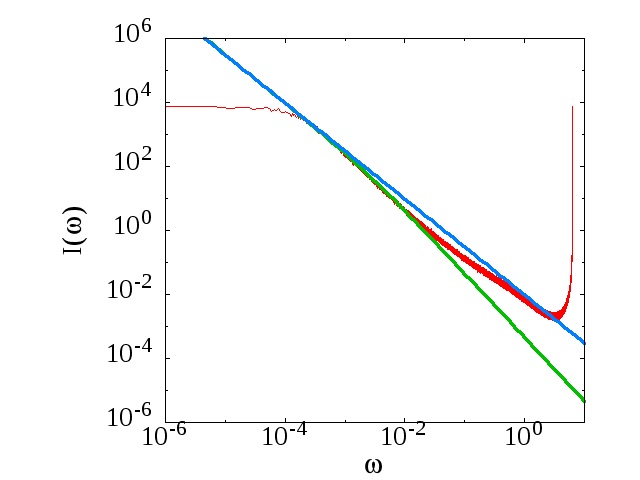} 
\label{f19}}
\subfigure[]{
\includegraphics[width=86mm]
{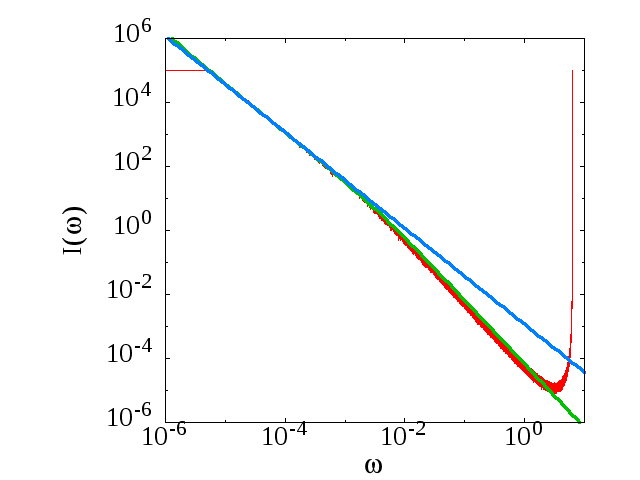}
\label{f20}}
\end{center}
\caption{Comparison between the power spectrum of $s_{L_x/2}$ 
and $I_E(\omega)$ for the system of size  $L_x=L_y=40$ and temperature (a) $T=2.0$ 
and (b) $T=0.5$.  In both figures,
red dots indicate the numerically obtained power spectrum of $s_{L_x/2}$,
green a little curved line shows $I_E(\omega)$ with  (a) $D=0.0158,L=29.9,L'=7.4,m_0=0.90$ and
(b) $D=0.000272$, $L=40.0,L'=0.8,m_0=1.00$, and blue straight line is a guide to the eyes for
the power law with exponent $-1.5$.
}
\label{f19and20}
\end{figure*}

Figures \ref{f19} and \ref{f20} show comparison between the power 
spectrum of $s_{L_x/2}$ 
and $I_E(\omega)$ at $T=2.0$ (Fig.~\ref{f19}) and $T=0.5$ (Fig.~\ref{f20}).
Compared with Fig.~\ref{f13}, 
we can see that $I_E(\omega)$ better fits the observed power spectrum than the
mere $-1.5$ power law does. Thus the time evolution of an interface in equilibrium Glauber dynamics is 
normal diffusion of an interface having width. The fit is better as temperature decreases. 
 Thus we have confirmed
that $I_E(\omega)$ reproduces the power spectrum of $s_{L_x/2}$ 
in frequency range $(\omega<0.01)$ at $T=2.0$ 
and that in a wider frequency range $(\omega<1.0)$ at $T=0.5$.
The magnitude of the power spectrum is larger at higher temperature. 
This is mainly because the diffusion constant is larger in higher temperature 
and the effective system size is smaller in higher temperature as discussed before.
Difference in the spontaneous magnetization is not significant except near
the critical temperature. Deviation seen in high-frequency range at $T=2.0$ is due to approximation in the
employed profile Eq.~(\ref{eq20}) and fluctuations around the mean profile in Fig.\/\ref{f18}.

\subsection{Equilibrium system with the KSC dynamics}
\begin{figure}[h]
 \centering
   \includegraphics[width=86mm,]{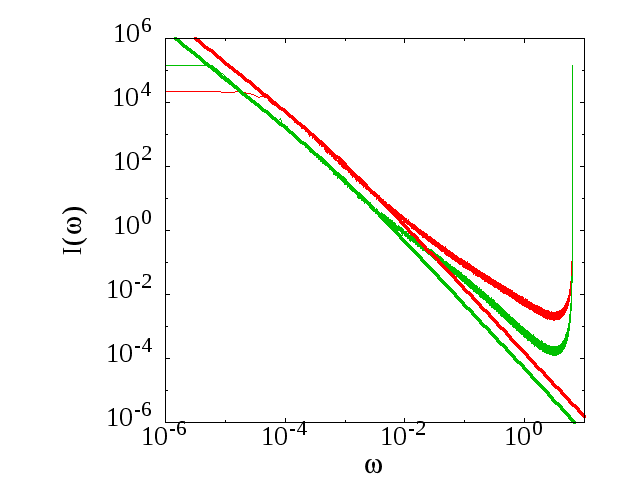}
   \caption{Comparison between the numerically obtained power spectrum of $s_{L_x/2}$ from
the KSC dynamics and $I_E(\omega)$ for the system of size $L_x=L_y=40$. 
Red (upper) and green (lower) dots show the power spectrum at $T=2.0$ and $1.0$, respectively. 
Red (upper) and green (lower) lines indicate $I_E(\omega)$ for $T=2.0$ ($D=0.00491, L=29.0, L'=7.3$) 
and $T=1.0$ ($D=0.000569, L=39.3, L'=2.4$), respectively.}
  \label{f21}
\end{figure}

In equilibrium systems, results from the KSC dynamics do not much differ from the Glauber 
case, though the diffusion constant is different from that for the Glauber dynamics at the 
same temperature. Figure \ref{f21} shows the comparison between the numerically obtained power spectrum
and Eq.~(\ref{eq29}). The low frequency region shows a good agreement between them.

\subsection{Nonequilibrium system}

We consider the case where the system is attached to two reservoirs 
at different temperatures $T_L$ and $T_R$. 
We assume that $T_L>T_R$ and states in the bulk of the system are 
updated by the KSC dynamics.  
As mentioned before, in nonequilibrium cases the stationary probability distribution 
of the interface position is exponential.
This is realized by adding a constant drift term to the Fokker-Planck equation as follows.
\begin{equation}
\frac{\partial P}{\partial t}=\frac{\partial}{\partial x}(-FP+D\frac{\partial P}{\partial x}), 
\label{eq39}
\end{equation}
where $F$ is a constant representing the force exerted on the interface.
The reflecting boundary condition in this case is 
\begin{equation}
\left(-FP+D\frac{\partial P}{\partial x}\right)\Big|_{x=0,L}=0. 
\label{eq40}
\end{equation}
From Eqs.~(\ref{eq39}) and (\ref{eq40}), we derive the stationary distribution as
\begin{equation}
P_{\mathrm{st}}(x)=\frac{Ke^{-KL}}{\sinh(K L)}\exp(2K x), 
\label{eq41}
\end{equation}
where $K=\frac{F}{2D}$.   
Thus we obtained an exponential distribution.
\begin{figure}[htbp]
 \centering
   \includegraphics[width=86mm]{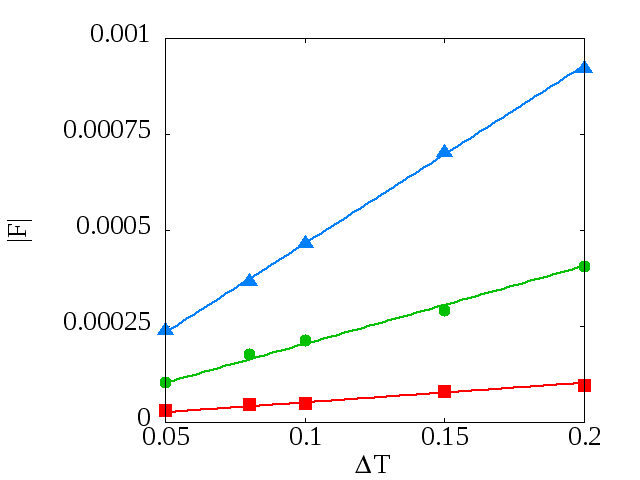}
   \caption{Drift force $F$  estimated from the probability distribution. 
The system size is $40\times40$ and the average temperature 
$T_{\mathrm{ave}}=1.0,1.5,2.0$. 
Red squares, green circles, and blue triangles indicate the results for $T_{\mathrm{ave}}=1.0,1.5$ and $2.0$, 
respectively. 
Red (lower), green (middle), and blue (upper) lines are fitting lines, $0.000512\Delta T, 0.00203\Delta T$, 
and $0.00464\Delta T$, respectively. 
To calculate the probability distribution to estimate the drift force $F$, 
$4.5\times10^7$ steps are taken after transient of $5\times10^6$ steps. 
} 
  \label{f22}
\end{figure}  

We examine whether Eq.~(\ref{eq39}) can also explain dynamical properties of the interface.
The power spectrum for the time sequence of the column-averaged magnetization at 
a horizontal position is derived in the same manner as before, which results in
\begin{widetext}
\begin{align}
 I_{NE}(\omega)&=  (2m_0)^2\frac{2}{DLL'^2}\frac{K L}{\sinh(K L)}
\left(\frac{\omega}{D}\right)^{-2}
\frac{\sqrt{\sqrt{(K^2)^2+(\frac{\omega}{D})^2}+K^2}}{\sqrt{(K^2)^2+(\frac{\omega}{D})^2}}
\left[-\frac{1}{\sqrt{2}}\cosh(K L')\right.
\nonumber\\
&+\exp\left(-L'\sqrt{\frac{\sqrt{(K^2)^2+(\frac{\omega}{D})^2}-K^2}{2}}\right) 
\left\{ 
\frac{1}{\sqrt{2}}
\cos\left(L'\sqrt{\frac{\sqrt{(K^2)^2+(\frac{\omega}{D})^2}-K^2}{2}}\right) \right.  
\nonumber \\
& \left.-\left(\frac{\omega}{D}\right)^{-1}
\left(\sqrt{(K^2)^2+(\frac{\omega}{D})^2}-K^2\right)
\sin\left(L'\sqrt{\frac{\sqrt{(K^2)^2+(\frac{\omega}{D})^2}-K^2}{2}}\right)\right\} 
\nonumber \\
&+\left.\left(\frac{\sqrt{\sqrt{(K^2)^2+(\frac{\omega}{D})^2}+K^2}}{\sqrt{(K^2)^2+(\frac{\omega}{D})^2}}\right)^{-1}L'\left(\frac{\sinh(K L')}{K L'}\right)\right].
\label{eq56}
\end{align}
\end{widetext}
Details of the derivation is given in Appendix A.

When $K \to 0$, it converges to $I_E(\omega)$.
If $|K| D\ll\omega$ and $|K| L'\ll1$, we have
\begin{equation}
I_{NE}(\omega)\sim \frac{K L}{\sinh(K L)}I_E(\omega). \label{eq55}
\end{equation}
\begin{figure}[htbp]
 \centering
   \includegraphics[width=86mm]{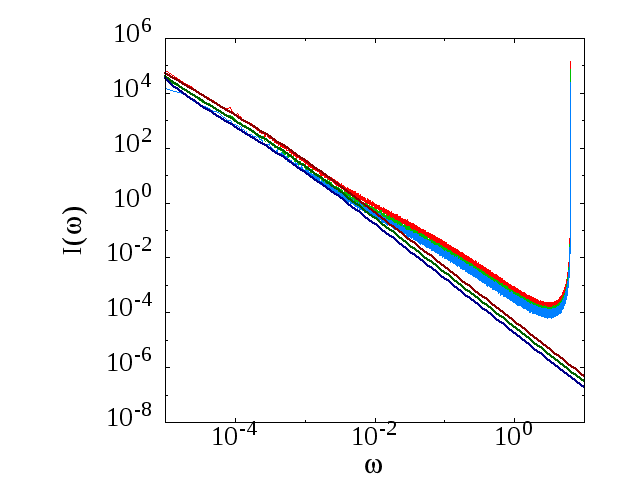}
   \caption{Comparison between the simulation results for the power spectra of 
$s_{L_x/2}$ ($T=1.0,L_x=L_y=40$) and $I_{NE}(x)$. Red (upper), green (middle), and blue (lower) dots indicate the numerical results for $\Delta T=0.0$, $0.1$ and 
$0.15$.  Dark red (upper), dark green (middle), and dark blue (lower) lines show corresponding 
$I_{NE}(x)$ for the same color.  Other parameter values are 
$D=0.000569$, $L=39.3$, $L'=2.4$, and $F=0.000512\Delta T$.
}
  \label{f23}
 \end{figure}
\begin{figure}[htbp]
 \centering
   \includegraphics[width=86mm]{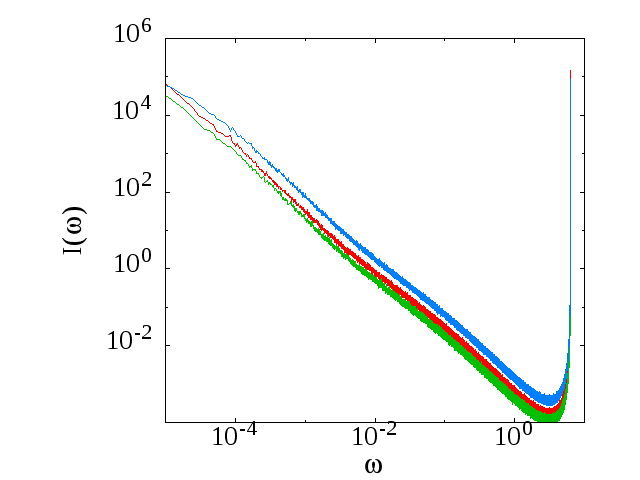}
   \caption{
Comparison between the power spectra of $s_{L_x/2}$ for the system of size $L_x=L_y=40$
in equilibrium [$T=1.0$, red (middle) dots] and 
in nonequilibrium [$T_{\mathrm{ave}}=1.0$, $\Delta T=0.1$, green (lower) dots)].
Blue (upper) dots represent the power spectrum of $s_{X_h}$ in nonequilibrium system 
($T=1.0$, $\Delta T=0.1$, $X_h=7$).
}
  \label{ft3}
 \end{figure}
\begin{figure}[htbp]
 \centering
   \includegraphics[width=86mm]{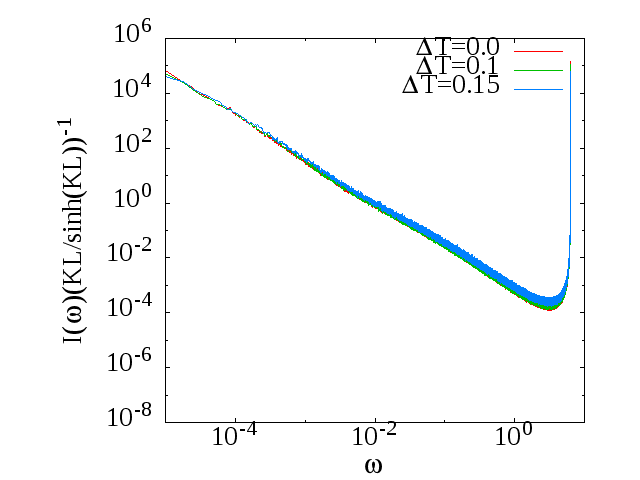}
   \caption{Scaling of the power spectra of $s_{L_x/2}$ ($T=1.0,L_y=40$). 
The vertical axis is $I(\omega)(\frac{K L}{\sinh(K L)})^{-1}$. 
They are scaled by $(\frac{K L}{\sinh(K L)})$.}
  \label{f24}
\end{figure}

Next, we compare the the numerical results with the obtained $I_{NE}(\omega)$. 
We use the same values for the  diffusion constant $D$, the effective length $L$, 
and the width $L'$ as in the  equilibrium system at temperature $T_{\mathrm{ave}}$.
Drift force $F$ is estimated from the probability distribution of the interface. 
Figure.~\ref{f22}  shows that the drift $F$ is proportional to the temperature difference 
$\Delta T$ as
\begin{equation}
|F|=f(T)\Delta T, \label{eq57}
\end{equation} 
where the constant $f(T)$ is estimated from Fig.~\ref{f22}.

Figure \ref{f23} shows the power spectra when $\Delta T$ is very small. 
We note that the larger the temperature difference $\Delta T$ is,
 the smaller the magnitude of the power spectrum is and that  
$I_{NE}(\omega)$ reproduces numerical results in a wide frequency range ($\omega<0.01$). 
Thus the motion of an interface is considered as normal diffusion with a constant drift to
higher temperature region. 
Addition of the drift term is a nonequilibrium effect.
Then, the interface prefers to stay in the high temperature region rather than the center. 
It causes reduction of the magnitude of the power spectrum for $s_{L_x/2}$ in 
nonequilibrium states.
On the other hand, if we consider the position $x_h$ defined by
$\int_0^{x_h}P_{st}(x')dx'=1/2$, the power spectrum of $s_{i}$ takes a maximum value
near $i\simeq x_h$. 
Let $x_h$ be rounded off to integer $X_h$.  
Figure \ref{ft3} shows the power spectrum of $s_{Lx/2}$ in equilibrium and nonequilibrium
states, and that of $s_{X_h}$ in nonequilibrium condition ($\Delta T=0.1$).
We see that the power spectrum of $s_{L_x/2}$ is larger in equilibrium than in 
nonequilibrium, but the power spectrum of $s_{X_h}$ surpasses the both of them.
Figure \ref{f24} shows power spectra with various $\Delta T$, 
which satisfy the scaling form Eq. (\ref{eq55}). 
In higher frequencies, deviation from $I_{NE}(\omega)$ is evident.
It is considered to be attributed to the approximation of the interface profile and fluctuations.

\section{Summary}

In this paper, we have studied interface motion in the two-dimensional Ising model
in equilibrium and noneqilibrium situations.  
To numerically simulate the systems effectively 
in low temperature, where Creutz dynamics freezes, 
we have devised the KSC dynamics by combining the Creutz dynamics and KS dynamics. 
Using the KSC dynamics, we have calculated thermal conductivity with and without
an interface and found that the thermal conductivity is larger in the system 
without an interface than that with an interface. 
The thermal conductivity is divided into the sum of contributions from Creutz 
and KS dynamics. 
Interestingly, the existence of an interface affects the two contributions in an opposite way.
When an interface exists, $\kappa_c(T)$ is larger and $\kappa_k(T)$ is smaller than
in the case of no interfaces.

Next, we investigated the probability distribution of the interface position in the KSC 
dynamics. We have found that the distribution is biased to the higher temperature 
region and that it is well approximated by an exponential distribution 
if temperature difference is very small.

In Sec. V, we have analyzed interface dynamics 
by using the power spectrum of a time sequence
of a spin at a site or column averaged magnetization.  
This is a generalization of the method developed in \cite{3} to two dimensions.  
In the previous paper \cite{3}, it was shown that 
the power spectrum of time sequence of a field variable at a position shows 
the power law behavior with exponent $-3/2$ due to the diffusive motion of
the interface.  Thus the most straightforward extension to two dimensions is
examining the power spectrum of a spin at a fixed site. However, we have seen that
in such a spectrum the power law is covered with large-frequency noise and obscure. 
Although the power spectrum of column averaged magnetization also deviates from the 
power law, we have revealed that the deviation is largely explained by 
taking the interface width into account. 
Moreover, the drift force in nonequilibrium dynamics affects the magnitudes 
of the power spectrum.  
Thus, in the two-dimensional Ising model the interface with a width
undergoes a diffusive motion with drift force to the higher temperature side
in the nonequilibrium situation.
Since the position of the interface is related to the magnetization.
it may be argued that the power spectrum of magnetization should have 
enough information for the interface motion. However, the motion of magnetization
is purely diffusive and the power spectrum shows only $\omega^{-2}$ behavior.
We cannot derive information of interface width from such a spectrum.

The method in \cite{3} and the present paper can be extended to the 
three-dimensional systems. Because the three-dimensional Ising model 
has the roughening transition \cite{10},
we may discuss roughening transition through the power spectrum of spin values.
It is a future problem.

\acknowledgments
The authors thank Satoshi Yukawa for useful discussions.

\appendix
\section{}
We start with Eq.~(\ref{eq39}) with the boundary condition (\ref{eq40}).
The transition probability is readily obtained as  
\begin{widetext}
\begin{eqnarray}
P(x,t|x_0,0)&=&\left(\frac{Ke^{-K L}}{\sinh(K L)}\right)^2\exp(2K(x+x_0))\\
          &  &+\sum_{n}\frac{2}{L}\exp(-k_nt)\frac{\exp(K(x+x_0))e^{-K L}}{\alpha_n^2+K^2}(\alpha_n\cos\alpha_nx_0+K\sin\alpha_nx_0)(\alpha_n\cos\alpha_nx+K\sin\alpha_nx), 
\label{eq45}
\end{eqnarray}
\end{widetext}
where $\alpha_n$ and $k_n$ are defined as 
\begin{equation}
\alpha_n=\frac{n\pi}{L} \quad (n=1,2,3,.....),
\label{eq46}
\end{equation}
and
\begin{equation}
k_n=D(\alpha^2_n+K^2) \quad (n=1,2,3,.....).
\label{eq47}
\end{equation}
We use Eq.~(\ref{eq20}) for the profile of the column-averaged magnetization with
an interface of width $L'$.  Then, the autocorrelation function of 
$\delta s_y(t)=s_y(t)-\langle s_y\rangle_{st}$ is obtained as
\begin{widetext}                                   
\begin{eqnarray}
\begin{aligned}
\lefteqn{\langle{\delta s_y(t)\delta s_y(0)}\rangle=}\\ 
&\sum_n\left(\frac{2}{L^2}\right){\left(\frac{2}{L'}\right)}^2\frac{e^{2K y}}{(\alpha_n^2+K)^3}\frac{K L}{\sinh(K L)}\exp(-k_n t)\{\frac{1+(-1)^{n+1}\cos({2\alpha_n y})}{2}(K\cos(\alpha_n\frac{L'}{2})\sin(K\frac{L'}{2})\\
&+\alpha_n\sin(\alpha_n\frac{L'}{2})\cos(K\frac{L'}{2}))^2+\frac{1+(-1)^n \cos(2\alpha y)}{2}(\alpha_n\cos(\alpha_n\frac{L'}{2})\sin(K\frac{L'}{2})-K\sin(\alpha_n\frac{L'}{2})\cos(K\frac{L'}{2}))^2 \\
&+2\sin(\alpha_ny)\cos(\alpha_n y)(K\cos(\alpha_n\frac{L'}{2})\sin(K\frac{L'}{2})+\alpha_n\sin(\alpha_n\frac{L'}{2})\cos(K\frac{L'}{2}))(\alpha_n\cos(\alpha_n\frac{L'}{2})\sin(K\frac{L'}{2})\\
&-K\sin(\alpha_n\frac{L'}{2})\cos(K\frac{L'}{2}))\}
\end{aligned}
\label{eq50}
\end{eqnarray} 
We put $y=L/2$ and  take Fourier transform to obtain the power spectrum
\begin{equation}
\begin{split}
I(\omega)&=\sum_{n=1}^{\infty}\frac{2k_n}{k_n^2+\omega ^2}
\frac{1}{L^2}\left(\frac{2}{L'}\right)^2\frac{K L}{\sinh(K L)}\frac{1}{(\alpha_n^2+K^2)^2}\{\sinh^2(K\frac{L'}{2})+\sin^2(\alpha_n\frac{L'}{2})\}. 
\end{split}
\label{eq51}
\end{equation}
When $L$ is large, the sum over $n$ can be replaced by the integral
\begin{equation}
\begin{split}
I(\omega)\sim\frac{1}{\pi L}\int_{0}^{\infty}dx \frac{2D}{D^2(x^2+K^2)^2+\omega ^2}\frac{K L}{\sinh(K L)}\frac{1}{(x^2+K^2)}\{\sinh^2(K\frac{L'}{2})+\sin^2(\frac{L'}{2}x)\} .
\end{split}
\label{eq52}
\end{equation}
Using residue calculus, we arrive at
\begin{equation}
\begin{aligned}
 I(\omega)\sim & \frac{2}{DLL'^2}\frac{K L}{\sinh(K L)}\left(\frac{\omega}{D}\right)^{-2}
\frac{
\sqrt{\sqrt{(K^2)^2+(\frac{\omega}{D})^2}+K^2}}
{\sqrt{(K^2)^2+(\frac{\omega}{D})^2}}
\left[{-\frac{1}{\sqrt{2}}}\cosh(K L')\right.\\
&+\exp\left(-L'
\sqrt{\frac{\sqrt{(K^2)^2+(\frac{\omega}{D})^2}-K^2}{2}}
\right)
\left\{
\frac{1}{\sqrt{2}}
\cos\left(L'\sqrt{\frac{ \sqrt{(K^2)^2+(\frac{\omega}{D})^2} -K^2 }{2}}\right) \right.  \\
& \left.
-\left(\frac{\omega}{D}\right)^{-1}
\left(\sqrt{(K^2)^2+(\frac{\omega}{D})^2}-K^2\right)
\sin\left(L'\sqrt{ \frac{\sqrt{(K^2)^2+(\frac{\omega}{D})^2}-K^2}{2} }\right)
\right\}
\\
& +\left.\left(\frac{\sqrt{\sqrt{(K^2)^2+(\frac{\omega}{D})^2}+K^2}}{\sqrt{(K^2)^2+(\frac{\omega}{D})^2}}\right)^{-1}L'\left(\frac{\sinh(K L')}{K L'}\right)\right].
\end{aligned}
\label{eq53}
\end{equation}
\end{widetext}

\end{document}